\documentstyle[12pt]{article}
\newcommand{\EQ}{\begin{equation}}
\newcommand{\EN}{\end{equation}}
\newcommand{\bea}{\begin{eqnarray}}
\newcommand{\eea}{\end{eqnarray}}

\newcommand{\be}{\beta}

\newcommand{\kb}{\bar{k}}
\newcommand{\ib}{\bar{\imath}}

\begin{document}
\setcounter{page}{0}
\topmargin 0pt
\oddsidemargin 5mm
\renewcommand{\thefootnote}{\fnsymbol{footnote}}
\newpage
\setcounter{page}{0}
\begin{titlepage}
\begin{flushright}
USP-IFQSC/TH/92-53
\end{flushright}
\vspace{0.5cm}
\begin{center}
{\large {\bf Couplings in  Affine Toda Field Theories}} \\
\vspace{1.8cm}
{\large Andreas Fring
\footnote{FRING@BR.ANSP.USP.IFQSC}}\\
\vspace{0.5cm}
{\em Universidade de S\~ao Paulo, \\
Caixa Postal 369, CEP 13560 S\~ao Carlos-SP, Brasil}\\
\vspace{3cm}
\renewcommand{\thefootnote}{\arabic{footnote}}
\setcounter{footnote}{0}
\begin{abstract}
{We present a systematic derivation for a general formula for the n-point
coupling constant valid for affine Toda theories related to any simple Lie
algebra {\bf g}. All n-point couplings with $n \geq 4$ are completely
determined
in terms of the masses and the three-point couplings. A general fusing rule,
formulated in the root space of the Lie algebra, is derived for all n-point
couplings. }
\end{abstract}
\vspace{.3cm}
\centerline{December 1992}
 \end{center}
\end{titlepage}
\newpage
\section{Introduction}

Affine Toda field theories describe $r$, being the rank of a particular Lie
 algebra {\bf g}, massive scalar fields in a relativistically invariant manner.
Due to their property to be integrable, they have  been subject of
investigation
for more than a decade \cite{MOP}, which has recently seen some revival since
it has been conjectured \cite{HMEY} that they might provide an explicit
Lagrangian version of integrable deformations of conformal field theories.
Seminal work on the latter approach has been carried out by Zamolodchikov
\cite{Zamo}, who argued that although the conformal symmetry is broken by
a perturbation, for certain cases the theory would still maintain its
integrability, i.e. posses an infinite number of integrals of motion as a
relic from the conformal field theory. The process of conformal symmetry
breaking in this context simply corresponds to an affinisation of the
Lie algebra {\bf g} underlying the conformally invariant Toda theories.
\par
Although it has turned out, from an application of the thermodynamic Bethe
Ansatz \cite{KM}, that only those theories with purely imaginary coupling
constant $\be$ correspond to perturbed minimal models \cite{BPZ}, whereas
those with $\be$ real reproduce a central charge of the Virasoro algebra
equalin
   g the
rank of the algebra, the latter theories remain a subject of interest in
their own right since they might give new insight into general principles
of field theory beyond perturbation theory. This hope is sustained by the
fact that it has been possible to understand the on-shell physics by making
reasonable Ans\"atze for scattering matrices [6-11,4]
which satisfy all consistency requirements demanded by general principles
of axiomatic quantum field theory (in particular the bootstrap equation),
 supported by perturbative checks \cite{ZZ,AFZ,BS}.
\par
For the simplest models of affine Toda field theories, i.e. the Sinh-Gordon
theory ($A_1$) and the Bullough-Dodd model ($A_2^{(2)}$) \cite{FMS}, it has
been possible to extend the knowledge off-shell and compute form factors. In
this approach the proper understanding of the n-point coupling is of vital
importance. Since it has turned out that many of the classical features
of affine Toda field theories, in particular for the simply laced case, survive
in the quantum version, it is desirable to obtain explicit values for the
n-point couplings. Generalizing the work in \cite{FLO} by extending it to
higher order, a systematic derivation of expressions for all n-point couplings
valid for any Lie algebra will be the subject of investigation in this paper.

\section{The n-point coupling constant}

Following the notation of \cite{FLO,FO} we shall start with the Lagrangian
density of affine Toda field theory corresponding to any simple Lie algebra
{\bf g}
\EQ
{\cal L} = tr \left( \frac{1}{2} \partial_{\mu} \Phi \partial^{\mu} \Phi -
e^{\beta \Phi} E e^{- \beta \Phi} E^{\dag} \right) \;\; , \label{eq: lagra}
\EN
proposed in this form originally in \cite{Freeman}. Here $\Phi$ denotes a field
depending on x,t and being an element of the Cartan subalgebra {\bf h}.
Taking the coupling constant $\beta$ to be real and using the fact that the
cyclic element $E$ \cite{Kostant} commutes with its hermitian conjugate
$E^{\dag}$, the theory will posses, on the contrary to $\beta$ being purely
imaginary \cite{HLOT}, a unique classical vacuum at $\Phi=0$ due to the
vanishing of the linear term in $\Phi$.
\par
For what follows the existence and properties of a second Cartan subalgebra
${\bf h'}$, furnished by the set of all elements commuting with $E$, will be
of vital importance. Taking two elements, say $T_i$ and $T_j$, which commute
with the cyclic element, the Jacobi identity involving this three elements
yields that $[ T_i, T_j ]$ commutes with $E$ too. Furthermore from
$tr(E [ T_i, T_j ] ) = tr ( T_j [ T_i, E ])$ one concludes that the commutator
of $T_i$ and $T_j$ vanishes and therefore ${\bf h'}$ is indeed a Cartan
subalgebra.
\par
Taking now $T_3 \in {\bf h}$ to be the generator of the principal $SU(2)$-
subalgebra embedded in {\bf g}, or equivalently the vector of which the
scalar product with any root gives its height, the so-called principal element
\EQ
S:= e^{ \frac{2 \pi i T_{3} }{h} }
\EN
can be defined, the adjoint action of which on ${\bf g}$ provides a
$Z\!\!\!\!Z^n$ grading
\EQ
{\bf g} = \bigoplus_{t=0}^{h-1} g_t   \qquad \qquad \qquad  S g_t S^{-1} =
\omega^n g_t   \;\; ,
\EN
with $\omega = \exp( \frac{2 \pi i}{h})$. The order of $S$ coincides with the
Coxeter number of {\bf g}. Denoting the generators in the
Cartan-Weyl basis of ${\bf h'}$ and {\bf h}, by $h_i$ and $H_i$,
respectively, the action of the principal element yields that $H_i \in
{\bf g}_0$ and $h_i \in {\bf g}_1$. On the other hand the adjoint action
of $S$ on the step-operators can be used to define a Coxeter transformation
$\sigma$, a product of Weyl reflections in some complete set of simple roots
$\alpha_i$, $S e_{\alpha} S^{-1} = e_{\sigma{(\alpha)}}$.
Since all Coxeter transformations are conjugate to each other, it is necessary
to make a particular choice. As has been first pointed out in \cite{Steinberg}
it is possible to select out two special elements of the Weyl group, $\sigma_+$
and $\sigma_-$, consisting out of products of Weyl reflections solely in
simple ``white" and ``black" roots, respectively. Here the two colours are
associated to the vertices of the Dynkin diagram of {\bf g} in such a way that
no two vertices of the same colour are connected. Selecting furthermore the
colour values $c(w)=-c(b)= 1$, the Coxeter transformation will separate the
set of all roots into $r$ distinct orbits $\Omega_i$, where each orbit is
represented by $\gamma_i = c(i) \alpha_i$. $\sigma_+$ and $\sigma_-$  can be
used to define unambiguously a  Coxeter element universally for all {\bf g},
i.e. $\sigma = \sigma_- \sigma_+$, for more details and notation refer
\cite{FLO,FO}. Using this facts, we shall now define the following element in
{\bf h}
\EQ
A_i := \frac{1}{\sqrt{h}} \sum_{n=1}^{h} S^n e_{\gamma_i} S^{-n} =
       \frac{1}{\sqrt{h}} \sum_{n=1}^{h}  e_{\sigma^n(\gamma_i)}
\EN
in which we choose to expand the field $\Phi$
\EQ
\Phi = \sum_{i=1}^{r} A_i \phi_i \;\; .
\EN
Here $\phi_i$ are the $r$, denoting the rank of the Lie algebra {\bf g},
scalar fields of the theory having the property $\phi_i^{*} = \phi_{
\bar{\imath}}$.

Expanding now the potential term in (\ref{eq: lagra}), the n$^{\rm th}$
order coupling  $C_{l_{1} \dots l_{n}}$  is defined as usual via the relation
\EQ
\frac{1}{n!} \; C_{l_{1} \dots l_{n}} \; \phi_{l_{1}} \dots \phi_{l{n}} \; .
\EN
For the Lagrangian (\ref{eq: lagra}) it turns out to be
\EQ
C_{l_{1} \dots l_{n}} = (-1)^{n+1} \be^n tr \left( E^{\dag} K_{l_{1} \dots
                        l_{n}}  \right) \label{eq: npoint}
\EN
where we introduced the tower of commutators
\EQ
K_{l_{1} \dots l_{n}} := \biggl[ \Bigl[ \bigl[ \dots [ E , A_{l_{1}} ] \dots
\bigr], A_{l_{n-1}} \Bigr], A_{l_{n}} \biggr] \;\; .
\EN
Or alternatively to (\ref{eq: npoint}), avoiding to compute $K$ to order
$n$, by exploiting the properties of the trace we obtain
\EQ
C_{l_{1} \dots l_{n}} = (-1)^{n+1} \be^n \left\{
\begin{array}{ll}
tr \left( K_{l_{1} \dots l_{\frac{n}{2} } } K^{\dag}_{\bar{l}_{{\frac{n}{2} +1}
} \dots \bar{ l}_{n} } \right) &  \quad {\rm for} \;\; n \;\; {\rm even}  \\
tr \left( [ K_{l_{1} \dots l_{\frac{n-1}{2} } }\; ,\;  K^{\dag}_{\bar{l}_{
{\frac{n+1}{2} } } \dots \bar{l}_{n-1}} ] A_{l_{n}} \right) &
\quad {\rm for} \;\; n \;\; {\rm odd} \;\; .  \label{eq: npointalt}
\end{array} \right.
\EN
We shall now attempt to compute the $K_{l_{1} \dots l_{n}}$ to all orders.
Since $E \in {\bf h'}$ we can expand it in terms of the basis of ${\bf h'}$,
i.e. $ E = q(1) \cdot h$ with $q(n)$ denoting the eigenvector of the Coxeter
element with eigenvalue $\omega^{s(n)}$
\EQ
q(n) = \sum_{k \in B} x_k(n) \alpha_k + e^{- \frac{i \pi}{h} s(n) }
       \sum_{k \in W} x_k(n) \alpha_k \;\; ,
\EN
$s(n)$ labeling the exponents of {\bf g} and $x_k(n)$ being the left
eigenvector of the Cartan matrix. We then compute with the usual commutation
relations in the Cartan-Weyl basis
\EQ
K_i = [E , A_i] = \frac{ q(1) \cdot \gamma_i}{\sqrt{h}} \sum_{n=1}^{h}
\omega^{-n} e_{ \sigma^{n}(\gamma_{i}) } \;\;  .
\EN
Evidently $K_i \in g_1 $, having furthermore the properties
\EQ
tr (K_i K_{\bar{\jmath}}^{\dag} ) = ( q(1) \cdot \gamma_i) ( q^{*}(1) \cdot
\gamma_{\bar{\jmath}} )\delta_{ij}  \qquad tr\left( K_i E \right) =
tr\left( K_i E^{\dag} \right) = 0
\EN
Next we compute recursively
\EQ
K_{ij} = \frac{q(1) \cdot \gamma_i}{h} \sum_{p,q =1}^h \omega^{-p}
S^p  \left[ e_{\gamma_{i}} ,  e_{\sigma^{q} (\gamma_{j} )} \right] S^{-p}
\;\;   .  \label{eq: seck}
\EN
There can only be two possibilities for this to be non-zero, that is
either the roots of the two step-operators add up to a third or to zero.
Starting with the former case, we know from the solution of the fusing
rule \cite{FO} that there exist only two possibilities for this to
happen, that is
\EQ
\left[ e_{\gamma_{i}} , e_{\sigma^{\zeta{(j)} - \zeta{(i)}} (\gamma_{j} )}
\right] = \varepsilon(i,j,\bar{k} )
e_{\sigma^{\zeta{\bar{(k)}} - \zeta{(i)}} (\gamma_{\bar{k}} )}
\EN
and a similar commutator after replacing $\varepsilon(i,j,\bar{k} )
\rightarrow \varepsilon'(i,j,\bar{k} )$ and $ \zeta(t) \rightarrow
\zeta'(t) = - \zeta(t) + \frac{ c(t) -1}{2} $ for $ t= i,j,k$.
Here the two triplets of integers $(\zeta(i),\zeta(j),\zeta(k))$ and
$(\zeta'(i),\zeta'(j),\zeta'(k))$ denote the two inequivalence classes
of solutions of the fusing rule for a three particle process.
{}From the fact that $A_i$ and $A_j$ commute we obtain that $\varepsilon$
and $\varepsilon'$ add up to zero and we can carry out the sum over $q$
in (\ref{eq: seck}), which after some re-arrangement yields
\bea
& & \frac{2 i}{\sqrt{h}} \sum_{\kb} \varepsilon(i,j,\kb) \frac{m_i}{m_j}\sin
\left( 2 \zeta(\kb) - 2 \zeta(i) + \frac{ c(i) - c(\kb)}{ 2} \right) K_{\kb}\\
\nonumber &=& \frac{1}{\beta} \sum_{\kb} \frac{c_{ij\kb}}{m_{k}^{2}} K_{\kb}
 \;\; . \\   \nonumber
\eea
The other possibility for the commutator in (\ref{eq: seck}) to be non-zero is
when $j \in \Omega_{\bar{\imath}}$ in which case the only contribution in the
sum over $q$ comes from $q = \frac{h}{2} + \frac{c(\bar{\imath}) - c(i) }{4}$.
By defining the quantity
\EQ
L_i := \frac{ q(1) \cdot \gamma_i }{h} \sum_{p=1}^{h} \omega^{-p} \sigma^p
(\gamma_i) \cdot h  \label{eq: baseh}
\EN
the total contribution in (\ref{eq: seck}) will be
\EQ
K_{ij} = \frac{1}{\beta} \sum_{\kb} \frac{c_{ij\kb}}{m_{k}^{2}} K_{\kb} + L_i
\delta_{\ib j}  \;\;  .
\EN
It turns out to be un-practical to carry on with this expression to higher
order or even to take the trace at this stage will not give an obvious
expression involving quantities which posses an obvious physical
interpretation. As so often in this context the solution lies in a change of
base.
\par
{}From the adjoint action of $S$ on $L_i$ we obtain that $L_i \in {\bf h'}$,
and
further with the fact that they are linearly independent and commute we obtain
that they furnish a basis of ${\bf h'}$. However taking the trace we observe
that it is not orthogonal. None-the-less, we can construct such a base by
means of a particular element in $ {\bf g}_{s(n)}$, defined as
\EQ
E_n := q(n) \cdot h \;\;  .
\EN
{}From the complex conjugation of the eigenvalue equation for $q(n)$ and the
equation for the inner product of $q(n)$ with $\gamma_i$, \cite{FO} equation
(3.7), we derive
\EQ
q^{*}(n) = q( r + 1 -n) \;\; .
\EN
and further from the unitarity of the Coxeter element we obtain
\EQ
q^{\dag}(n) q(m) = \delta_{nm}  \;\;  .
\EN
Therefore
\EQ
E_n^{\dag} = E_{r+1-n} \qquad \hbox{and} \qquad tr \left( E^{\dag}_n E_m
\right)= \delta_{nm} \;\; ,
\EN
Hence we have  obtained a base of ${\bf h'}$ with the property $tr(h_i h_j) =
\delta_{ij}$
\EQ
h_i = \sum_{n=1}^{r} q_i^{*}(n) E_n \;\; .
\EN
Thus on substituting this into (\ref{eq: baseh}) we obtain
$$
L_i = i q(1) \cdot \gamma_i \sum_{n=1}^{r} \alpha_i^2 x_i(r+1-n) \sin
\left( \frac{2\pi }{h} \right) e^{-i s(r+1-n) \frac{1 - c(i)}{2}} \sum_{p=1}^{
h} e^{ -\frac{2 \pi i}{h} ( s(r+1-n) + 1) p } E_n  \; .
$$
The sum over p will always vanish except when $n=1$ and therefore this
becomes
\EQ
i q(1) \cdot \gamma_i \alpha_i^2 x_i(r) \sin \frac{2 \pi}{h}
e^{ -i \frac{ 2 \pi i s(r)}{h} \frac{ 1 - c(i) } {2} }  E = \frac{m_{i}^{2}}
{\beta^{2}} E
\EN
an  so we can re-write (\ref{eq: seck}) in a more suitable way as
\EQ
K_{ij} = \frac{1}{\beta} \sum_{\kb} \frac{C_{ij\kb}}{m_{k}^{2}} K_{\bar k} +
\frac{m_{i}^{2}}{\beta^{2}} E \delta_{\ib j}  \;\;  . \label{eq: kzwei}
\EN
Thus further K's can now be obtained recursively
\bea
K_{ijk} &=& \frac{1}{\be^2} \sum_{\bar{l},\bar{m}} \frac{C_{ij\bar{l}}
C_{\bar{l}k\bar{n}}}{m_l^2 m_n^2} K_{\bar{n}}  + \frac{m_i^2}{\be^2}
\delta_{ \ib j} K_k+\frac{C_{ijk}}{\be^3} E  \\
K_{ijkl} &=& \frac{1}{\be^3} \sum_{\bar{t},\bar{n},\bar{s}} \frac{C_{ij\bar{t}}
C_{\bar{t}k\bar{n}}C_{\bar{n}l\bar{s}}} {m_t^2 m_n^2 m_s^2} K_{\bar{s}} +
\frac{1}{\be^3} m_i^2 \delta_{\ib j} \sum_{\bar{t}}
\frac{C_{kl\bar{t}}}{m_t^2} K_{\bar{t}}  + \frac{C_{ijk}}{\be^3} K_l \\
& & \frac{1}{\be^4} \left( \sum_{\bar{t}} \frac{C_{ij\bar{t}}C_{\bar{t}kl}}
{m_t^2} + m_i^2 \delta_{\ib j} m_k^2 \delta_{\kb l} \right)  E  \;\; .\nonumber
\eea
{}From the commutation relations it is now evident how this generalizes and we
find the following Feynman like rules for the construction of a general
$K_{l_{1} \dots l_{n}}$: Start with the contraction of some indices always
from the left hand side. A fusing of two indices to a third, say ij to k,
contributes a factor
$$  \frac{1}{\be}  \sum_{\kb} \frac{C_{ij \kb }}{ m_k^2} \;\;  ,  $$
whereas the annihilation of two indices, i and j say, contributes the
mass matrix
$$  \frac{m_i^2 \delta_{\ib j} }{\be^2}   \;\;   .$$
If at the end of the contraction process there is still an index left it
will contribute a $K_i$, otherwise if all indices are consumed we close
with an $E$. Performing all possible contractions then gives the final answer
for $K_{ l_{1} \dots \l_{n} }$.
\par
The proof of this rule follows by induction. Clearly, it is true for
$K_{ l_{1} l_{2} }$, since we can either contract the two indices to a third
or annihilate them, and thus reproduce equation (\ref{eq: kzwei}). Assuming
the validity of the rule for $K_{ l_{1} \dots \l_{n-1} }$ we can employ the
fact that $ K_{ l_{1} \dots \l_{n} } = \left[ K_{ l_{1} \dots \l_{n-1} } ,
A_{l_{n}} \right]$ and deduce the expression for $K_{ l_{1} \dots \l_{n} }$.
According to the rule the terms in $K_{ l_{1} \dots \l_{n-1} }$ can only
finish with $K_i$ or $E$, corresponding to the situation in which the last
index is left or consumed, respectively. In the former case the additional
index $l_n$ can either be contracted to a third or annihilated, which
corresponds to an extension of this terms by $\left[ K_i, A_{l_{n}}\right]
= K_{i l_{n}}$ . Whereas in the latter case there is no index left and we
have to extend the terms with $K_{l_{n}}= \left[ E ,A_{l_{n}} \right]$
\par
According to (\ref{eq: npoint}) or (\ref{eq: npointalt}) we can now compute
the n-point coupling constant. We recover correctly the mass matrix $C_{ij}
= - m_i^2 \delta_{\ib j}$ the three point coupling and further
\bea
C_{ijkl} &=& - \sum_{\bar{t}} \frac{C_{ij\bar{t}} C_{\bar{t} kl}}{m_t^2} -
m_i^2 \delta_{\ib j} m_k^2 \delta_{\kb l}   \label{eq: bcds}  \\
C_{ijkln} &=& \sum_{\bar{t},\bar{u}} \frac{C_{ij\bar{t}} C_{\bar{t}k\bar{u}}
C_{\bar{u}ln}}{m_t^2 m_u^2} + m_i^2 \delta_{\ib j} C_{kln} + C_{ijk} m_l^2
\delta_{\bar{l} n}
\eea
Equation (\ref{eq: bcds})  confirms the result of \cite{CM,BS} derived from
the assumption that certain projection operators exist.
\par
So we can cast the formula for the $n$-point coupling into the following
compact form
\EQ
C_{l_{1} \dots l_{n}} =  (-1)^{n+1} \sum_{t=1}^{n-2} \sum\limits_{
\leftrightarrow \atop x}
\frac{x_1 \dots x_t}{{\cal N}_t} m_{l_{n-1}}^2 \delta_{\bar{l}_{n-1}  l_{n}}
\label{eq: closedc}
\EN
where the $x_i$ for each particular $t$ are given by
\EQ
x_1, \dots, x_{2(t+1) -n} = \sum_{\kb} \frac{C_{\mu \nu \kb}}{m_k^2} \qquad
x_{2t +3 -n}, \dots, x_t = m_{l{\nu}}^2 \delta_{\bar{l}_{\nu}  l_{\mu}} \;\; .
\EN
Here $\sum\limits_{\leftrightarrow \atop x}$ denotes the sum over all possible
permutations of the $x_i$. The factor ${\cal N}_t$ takes care of the
overcountin
   g
when symmetric terms are permuted. Its explicit value is given by
\EQ {\cal N}_t = (2t +2 -n) ! (n - t -2) !  \;\;  . \EN
Having carried out the sums in (\ref{eq: closedc}), the greek indices $\mu,
\nu, \dots $ have to be substituted in the same order as on  the left hand
side of this equation. A three-point coupling will always be connected to its
neighbouring term on the right via a sum over a dummy variable and a ``zero
momentum propagator".
\par
This formula follows simply from the fact that $ \frac{1}{\be^n}
C_{l_{1} \dots \l_{n} } $ corresponds to the term in front of $E$ in
$ K_{ l_{1} \dots l_{n} }$. Since the last two indices are always annihilated
in this term by construction and all the other terms are symmetric in the
$C$'s and $\delta$'s, up to a permutation of the $C$'s and $\delta$'s
among each other we obtain (\ref{eq: closedc}).

\section{Fusing Rules}
It appears now natural to pose the question of how this rules can be formulated
in the rootspace of the Lie algebra, that is how the fusing rule known for the
three-point coupling \cite{PD,FLO,FO} generalizes to higher order. In the same
way as in \cite{FLO} we check what is implied for the root space, from a
non-vanishing term in the coupling constant resulting from some non-zero
commut
   ator.  Starting at the lowest order, the second term in (\ref{eq: kzwei})
is only non-zero, and therefore $C_{ij}$, if there exist two roots in the
orbits $\Omega_i$ and $\Omega_j$ which add up to zero
\EQ
\sigma^{\zeta(i)} \gamma_i +  \sigma^{\zeta{(j)}} \gamma_j  = 0  \;\;  .
\EN
This can only be true if particle $j$ is the anti-particle of i, i.e. $j=\ib$,
in which case the mass matrix will be diagonal. The three-point coupling
 constant $C_{ijk}$ is known to be non-vanishing if there are three orbits
 $\Omega_i$, $\Omega_j$ and $\Omega_k$ such that
\EQ
\sigma^{\zeta(i)} \gamma_i + \sigma^{\zeta(j)} \gamma_j +
\sigma^{\zeta(k)} \gamma_k   = 0  \;\;  .
\EN
{}From the previous section we obtain that $C_{l_{1} \dots l_{n} }$ for
$ n \geq 4$ is non-zero if and only if there exist n roots in $\Omega_i$
for $i=1, \dots, n$ which add up to zero
\EQ
\sum_{i=1}^n \sigma^{\zeta(i)} \gamma_i = 0 \;\;  , \label{eq: ngon}
\EN
together with the additional constraint that the following $n-2$ triangles
exist
\bea
\sigma^{\zeta(1)} \gamma_1 + \sigma^{\zeta(2)} \gamma_2 &= & \sigma^{\zeta(
                   \bar{t}_{1}) } \gamma_{\bar{t}_{1}}  \nonumber \\
\sigma^{\zeta(\bar{t}_{1})} \gamma_{\bar{t}_{1}} + \sigma^{\zeta(3)} \gamma_3
   &=& \sigma^{\zeta(\bar{t}_{2}) } \gamma_{\bar{t}_{2}} \nonumber \\
            & \vdots  &    \label{eq: Dreiecke}       \\
\sigma^{\zeta(\bar{t}_{n-4})} \gamma_{\bar{t}_{n-4}} + \sigma^{\zeta(n-2)}
    \gamma_{n-2} &=& \sigma^{\zeta(\bar{t}_{n-3}) } \gamma_{\bar{t}_{n-3}}
                \nonumber \\
\sigma^{\zeta(\bar{t}_{n-3})} \gamma_{\bar{t}_{n-3}} + \sigma^{\zeta(n-1)}
    \gamma_{n-1} &=& \sigma^{\zeta(n) } \gamma_{n}
\;\;  .                \nonumber
\eea
Here the intermediate particles $\gamma_{\bar{t}_{i}}$ are elements of
$\Omega_i$ for $i= 0,1, \dots r$. By $\Omega_0$ we denote the ``orbit"
just consisting out of the zero vector, which means it is allowed that some
of the triangles collapse. In the same fashion as in \cite{FO} we can show
that each of the equations (\ref{eq: Dreiecke})
possesses two inequivalent solutions of equivalence classes, such that for
$C_{l_{1} \dots l_{n} }$ there will be $2n -4 $ solutions. We illustrate
the four possible solutions for a particular $\bar{t}$ of the fusing rule
giving rise to the first term in (\ref{eq: kzwei}) in figure 1. Setting
$\bar{t} \rightarrow 0$ will give rise to the additional term involving
the masses.
\par
We can make use as well of the relation between roots and weights, i.e.
$\gamma_i = ( \sigma_- - \sigma_+) \lambda_i$ and reformulate the whole set of
equations in the weight space, which might be used to find alternative
 formulations in terms of representations analogously to \cite{HB}.
\par
Taking the inner product of the whole set of vectors with $q(1)$ and projecting
equations (\ref{eq: ngon}) - (\ref{eq: Dreiecke}) into the velocity plane
will lead to n-gons, bounded by the masses $m_1, \dots , m_n$ of the fusing
particles. The further condition on this is that it has to be possible to
triangulate the n-gon by using as secants only the masses present in the
theory.
We illustrate this for the nine-gon in figure 2 for $C_{ijklnopqr}$. Notice the
possibility that some of the vertices might superimpose in case some of the
inner secants collapse.
\section{Conclusions}
Starting from the affine Toda field theory Lagrangian density we presented a
sys
   tematic derivation for the n-point coupling constant valid for any simple
Lie
    algebra {\bf g}. The masses and the three-point couplings turn out to be
the
fundamental quantities in which all the higher couplings can be completely
expressed. Fusing rules which provide a selection rule for non-vanishing
n-point couplings have been formulated in the root space, where again the
fundamental identities are the fusing rules for the non-vanishing three-point
coupling, i.e. that is the "triangle rule". Using the result that the classical
fusing rule for a non-vanishing three-point coupling becomes quantum
mechanicall
   y
equivalent to the bootstrap equation \cite{FO}, we might expect this to
generali
   ze. However, provided the analogy holds a ``bootstrap equation" involving
more than three particles will only contain redundant information what the
on-shell physics concerns and ``everything" can be extracted from the
three-particle bootstrap. However, off-shell the information about the
possible n-particle interactions will be of importance and the classical
information may be used to formulate a general theory of form factors for
affine Toda theories.
\par

{\em Acknowledgement}: I am  grateful to Roland K\"oberle, Giuseppe Mussardo
and
 David Olive for helpful discussions. Furthermore I gratefully acknowledge the
 support of the Funda\c{c}\~ao de Amparo \`a Pesquisa do Estado de S\~ao Paulo
 (FAPESP).

\newpage

 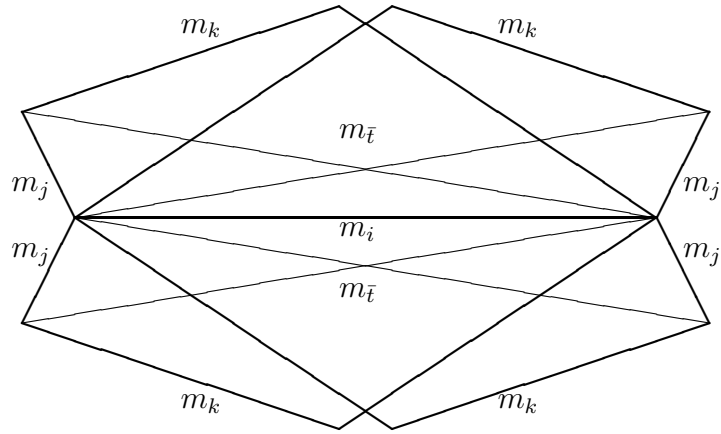
\begin{figure}
\begin{picture}(40,60)(60,470)
\thicklines
\put(220,560){\line( 1, 0){220}}
\put(220,560){\line( 3,-2){120}}
\put(220,560){\line( 3,2){120}}
\put(340,640){\line(3,-1){120}}
\put(440,560){\line( 1,2){20}}
\put(340,480){\line( 3, 1){120}}
\put(460,520){\line(-1, 2){ 20}}
\put(220,560){\line(-1,2){20}}
\put(220,560){\line(-1,-2){20}}
\put(440,560){\line(-3,2){120}}
\put(440,560){\line(-3,-2){120}}
 \put(200,600){\line(3,1){120}}
 \put(200,520){\line(3,-1){120}}
 \thinlines
\put(220,560){\line( 6 , 1 ){240}}
\put(220,560){\line( 6 , -1 ){240}}
\put(440,560){\line( -6 , 1 ){240}}
\put(440,560){\line( -6 , -1 ){240}}
\put(320,553){$m_i$}
\put(196,570){$m_j$}
\put(196,545){$m_j$}
\put(450,570){$m_j$}
\put(450,545){$m_j$}
\put(320,530){$m_{\bar{t}}$}
\put(320,590){$m_{\bar{t}}$}
\put(260,630){$m_k$}
\put(260,488){$m_k$}
\put(380,630){$m_k$}
\put(380,488){$m_k$}
\end{picture}
 \caption{Four-gons in the velocity plane for $C_{ijkl}$}
 \end{figure}
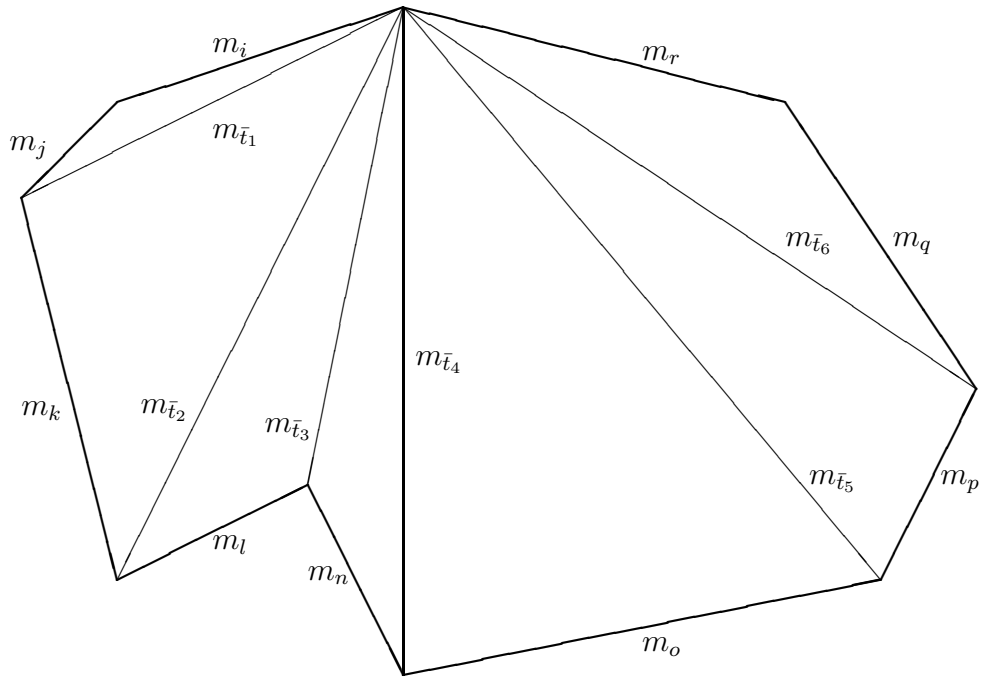
\begin{figure}
\setlength{\unitlength}{0.0125in}
\begin{picture}(40,180)(120,420)
\thicklines
\put(330,700){\line( -3, -1){120}}
\put(210,660){\line( -1, -1){40}}
\put(170,620){\line( 1, -4){40}}
\put(210,460){\line( 2, 1){80}}
\put(290,500){\line( 1, -2){40}}
\put(330,420){\line( 5, 1){200}}
\put(530,460){\line( 1, 2){40}}
\put(570,540){\line( -2, 3){80}}
\put(330,700){\line( 4, -1){160}}
\thinlines
\put(330,700){\line( -2, -1){160}}
\put(330,700){\line( -1, -2){120}}
\put(330,700){\line( -1, -5){40}}
\put(330,700){\line( 0, -1){280}}
\put(330,700){\line( 5, -6){200}}
\put(330,700){\line( 3, -2){240}}
\put(250,681){$m_i$}
\put(165,640){$m_j$}
\put(170,529){$m_k$}
\put(250,473){$m_l$}
\put(290,460){$m_n$}
\put(430,430){$m_o$}
\put(555,500){$m_p$}
\put(535,600){$m_q$}
\put(430,678){$m_r$}
\put(250,645){$m_{\bar{t}_{1}}$}
\put(220,530){$m_{\bar{t}_{2}}$}
\put(272,522){$m_{\bar{t}_{3}}$}
\put(335,550){$m_{\bar{t}_{4}}$}
\put(500,500){$m_{\bar{t}_{5}}$}
\put(490,600){$m_{\bar{t}_{6}}$}
\end{picture}
\caption{Triangulated nine-gon in the velocity plane for $C_{ijklnopqr}$}
 \end{figure}

\begin{thebibliography}{99}

\bibitem{MOP} A.V. Mikhailov, M.A. Olshanetsky and A.M. Perelomov,
{\em Comm. Math. Phys.} {\bf 79} (1981), 473; G. Wilson, {\em Ergod. Th.
Dyn. Syst.} {\bf 1} (1981) 361; D.I. Olive and  N. Turok, {\em Nucl. Phys.}
{\bf B257} [FS14] (1985) 277.
\bibitem{HMEY} T.J. Hollowood and  P. Mansfield, {\em Phys. Lett.} {\bf B226}
(1989) 73; T. Eguchi and  S.-K. Yang {\em Phys. Lett.} {\bf B224} (1989) 373.
\bibitem{Zamo} A.B. Zamolodchikov, Int. J. Mod. Phys. A1 (1989) 4235.
\bibitem{KM} T.R. Klassen and E. Melzer, {\em Nucl. Phys.} {\bf B338} (1990)
485.
\bibitem{BPZ} A.A. Belavin, A.M. Polyakov and A.B. Zamolodchikov, {\em Nucl.
 Phys.} {\bf B241} (1984) 333.
\bibitem{AFZ} A.E. Arinshtein, V.A. Fateev and A.B. Zamolodchikov,
{\em Phys. Lett.} {\bf 87B} (1979) 389.
\bibitem{KS} R. K\"oberle and J.A. Swieca, {\em Phys. Lett.} {\bf 86B}
(1979) 209; A.B. Zamolodchikov, {\em Int. J. Mod. Phys.} {\bf A3} (1988)
743; V. A. Fateev and A.B. Zamolodchikov, {\em Int. J. Mod. Phys.} {\bf A5}
(1990) 1025.
\bibitem{BCDS} H. W. Braden, E. Corrigan, P. E. Dorey and R. Sasaki, {\em
Phys. Lett.} {\bf B227} (1989) 411; H. W. Braden, E. Corrigan, P. E. Dorey
and R. Sasaki, {\em Nucl. Phys.} {\bf B338} (1990) 689.
\bibitem{CM} P. Christe and G. Mussardo, {\em Nucl.Phys.} {\bf B330} (1990)
465; P. Christe and G. Mussardo {\em Int. J. Mod. Phys.} {\bf A5} (1990) 1025.
    \bibitem{PD} P.E. Dorey, {\em Nucl. Phys.} {\bf B358} (1991) 654; P.E.
Dorey
   ,
 {\em Nucl. Phys.} {\bf B374} (1992) 741.
\bibitem{FO} A. Fring and D.I. Olive, {\em Nucl. Phys.} {\bf B379} (1992) 429.
\bibitem{ZZ} A.B. Zamolodchikov and Al. B. Zamolodchikov, {\em Ann. Phys.}
{\bf 120} (1979) 253.
\bibitem{BS} H.W. Braden and R. Sasaki, {\em Phys. Lett.} {\bf B255} (1991)
343; H.W. Braden and R. Sasaki, {\em Nucl. Phys.} {\bf B379} 377.
\bibitem{FMS} A. Fring, G. Mussardo and P. Simonetti, {\em Form Factors }
{\em for Integrable} {\em Lagrangian Field Theories, the Sinh-Gordon Model},
ISA
   S/92-146,
Imperial/TP/91-92/31,  to appear Nucl. Phys. B; A. Fring, G. Mussardo and P.
Simonetti, {\em Form Factors of the Elementary Field in the Bullough-Dodd
Model}, ISAS/EP/92/208, USP-IFQSC/TH/92-51; A. Fring, G. Mussardo and P.
Simonetti, {\em in preparation}.
\bibitem{FLO} A. Fring, H.C. Liao and D.I. Olive, {\em Phys. Lett.} {\bf B266}
(1991) 82.
\bibitem{Freeman} M. Freeman, {\em Phys. Lett.} {\bf B261} (1991) 57.
\bibitem{Kostant} B. Kostant, {\em Amer. J. Math} {\bf 81} (1959) 973.
\bibitem{HLOT} T. Hollowood, {\em Nucl. Phys.} {\bf B384} (1992) 523;
H.C. Liao, D.I. Olive, N. Turok, {\em Topological
Solitons in $A_r$ Affine Toda Theory} Imperial preprint TP/91-92/34; D.I.
Olive,
N. Turok, J.W.R. Underwood, {\em Solitons and the Energy-Momentum Tensor for
Affine Toda Theory} Imperial preprint TP/91-92/35 ; H. Aratyn, C.P.
Constantinidis, L.A. Ferreira, J.F. Gomes and A.H. Zimerman, {\em Hirota`s
Solitons in the Affine and the Conformal Affine Toda Models} IFT preprint
 IFT-P/052/92.
\bibitem{Steinberg} R. Steinberg, {\em Trans. Amer. Soc.} {\bf 91} (1959) 493.
\bibitem{HB} H.W. Braden, {\em J. Phys.} {\bf A25} (1992) L15.
\end{thebibliography}
\end{document}